\documentclass[
reprint,
amsmath,
amssymb,
aps,
longbibliography,
floatfix
]{revtex4-2}

\usepackage{graphicx}
\usepackage{xcolor}
\usepackage[english]{babel}
\usepackage[T1]{fontenc}
\usepackage[utf8]{inputenc}
\usepackage{dcolumn}
\usepackage{bm}
\usepackage{hyperref}
\usepackage[capitalise]{cleveref}
\usepackage[mathlines]{lineno}
\usepackage{multirow}
\usepackage{braket}

\usepackage{MnSymbol}
\usepackage{physics}

\begin{document}

\preprint{APS/123-QED}

\title{
Coherent Control of Quantum and Classical Correlations in Photoionization
}

\author{Axel Stenquist
}
\author{Jan Marcus Dahlström$^{{\text{\dagger}}}$
}

\affiliation{Department of Physics, Lund University, 22100 Lund, Sweden.}

\begin{abstract}
\noindent 

The ability to control quantum correlations in strongly driven systems is a central challenge across quantum science, with implications for ultrafast dynamics, quantum control, and information processing. In photoionization, the emitted electron and residual ion may form an entangled system whose correlations encode the underlying light–matter interaction, yet control of their generation and observable manifestation in continuum systems remains largely unexplored. Here we demonstrate phase-resolved control of electron–ion correlations using phase-locked pulse sequences in the strong-coupling regime. We show that entanglement can be halted and reshaped with attosecond precision, and that phase-dependent correlations can be redistributed into population-based correlations, leading to entanglement that is directly reflected in joint observables. These results establish a route to coherently shape entanglement in photoionization and open new  possibilities for accessing and controlling quantum correlations in systems where measurements are intrinsically basis constrained.
\\\\
\textbf{Keywords:} 
Coherent Control,
Correlation,  
Photoionization, 
Strong Coupling, 
Quantum Entanglement, 
Attosecond Physics
\end{abstract}

\maketitle

\begin{table}[b!]
\begin{flushleft}
  $^\text{\dagger}$ marcus.dahlstrom@fysik.lu.se
\end{flushleft}
\end{table}

\begin{figure*}
    \centering
    \includegraphics[width=1\linewidth]{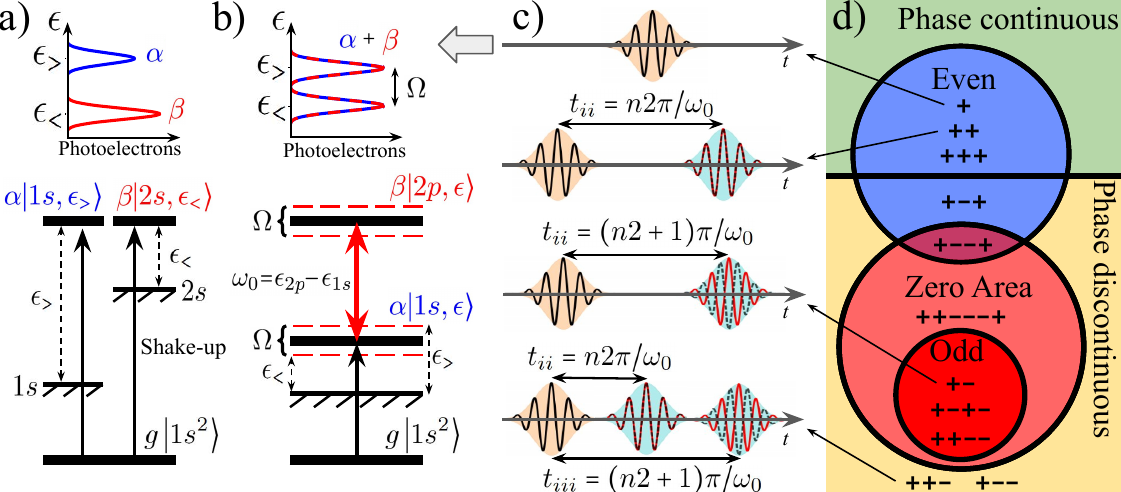}
    \caption{\textit{Energy level schematics and pulse sequence symmetries.} 
    a) Shake-up process, where the system is ionized to two non-degenerate states, yielding non-overlapping photoelectron distributions (top). 
    b) Photoionization, followed by strong coupling of the ion, yielding overlapping photoelectron distributions (top). 
    c) Various phase-locked pulse sequences used for quantum control, where $I = \{i,ii,iii,iv,...\}$ denotes the pulse number.
    d) Venn diagram categorizing pulse sequences, where $+$ and $-$ denote pulses which are in or out of phase with the first pulse, respectively. Pulse sequence properties include the zero-area condition and symmetries in time.}
    \label{fig:2a}
\end{figure*}

\section{Introduction}\label{sec:introduction}

Photoionization provides a natural platform for generating and probing quantum correlations on attosecond time scales, where an emitted electron and the residual ion form an entangled bipartite system. These correlations encode both the coherence of the driving field and decoherence of the underlying light–matter interaction, making them a sensitive probe of ultrafast dynamics \cite{pabst_decoherence_2011}. 
Advances in strong-field and attosecond physics have enabled time-resolved studies of the purity in photoionization, from atomic \cite{busto_probing_2022,mehmood_ionic_2023,ishikawa_control_2023} and molecular systems  \cite{vrakking_control_2021,koll_experimental_2022,ruberti_quantum_2022,shobeiry_emission_2024}, with entanglement being linked to strong coupling \cite{nandi_generation_2024} and attosecond delays in photoionization \cite{jiang_time_2024,makos_entanglement_2025}. Access to the reduced states of ions has been achieved using transient absorption \cite{goulielmakis_real-time_2010} and of electrons using laser-assisted photoionization \cite{laurell_measuring_2025,berkane_complete_2025}. Recently, a Bell test in ultrafast photoionization have been proposed with pairs of noncommuting operators \cite{ruberti_bell_2024}.

For an isolated electron–ion system described by a pure state $\ket{\Psi_{\text{e}^-\, \text{ion}}}$, the entanglement, $\cal E$, quantum correlation, $\mathcal{Q}$, and maximal classical correlation, ${\cal C}_\text{max}$ are quantified by the {\it von Neumann entropy} of the reduced density matrix,
\begin{equation}\label{Eq:correlation_pure}
    {\cal E} = {\cal Q} = {\cal C}_\text{max} = S_\text{vN}(\tilde \rho),
\end{equation}
where $\tilde \rho$ denotes the reduced states of ion or electron \cite{henderson_classical_2001,maziero_classical_2009,haroche_exploring_2006,cruz-rodriguez_quantum_2024}. This equivalence follows from the Schmidt decomposition, which establishes one-to-one correlations between electron and ion states, $S_\text{vN}(\tilde \rho_{\text{e}^-}) = S_\text{vN}(\tilde \rho_{\text{ion}})$. For ultrafast processes, where the joint system is well approximated as pure, this provides a natural framework for quantifying entanglement. 
However, a central challenge in complex quantum systems is that entangled states are defined in an abstract, optimal (Schmidt) basis, while experimentally accessible observables may be constrained to specific physical bases \cite{tichy_essential_2011}. Understanding how entanglement manifests under such constraints is therefore essential for connecting theory with measurement. 

In realistic settings, interactions with the environment modify these correlations. Spatial separation leads to decoherence and the emergence of preferred pointer states \cite{zurek_decoherence_2003} and noise decreases entanglement over time \cite{yu_sudden_2009}. Furthermore, spontaneous decay transfers entanglement to additional degrees of freedom \cite{stenquist_entanglement_2025}. Much of these complications are mitigated with experiments performed on ultrafast timescales. 
%
Here, the relative phases of photoelectron wave packets can be tuned with attosecond precision using pump-probe schemes. This raises the possibility of controlling not only the magnitude of entanglement, but also its structure and its manifestation in experimentally accessible observables.

In this work, we demonstrate phase-resolved control of electron–ion correlations in the strong-coupling regime using a sequence of coherent pulses. We show that entanglement generation can be halted and its observable structure can be manipulated independently of its total magnitude. By tracking the time-dependent buildup of correlations, we identify regimes where phase-dependent correlations are redistributed into population-based correlations that directly determine measurable quantities.
To describe this behavior, we introduce a basis-resolved characterization, termed {\it amplitude entanglement}, which isolates the contribution of population structure from phase-dependent interference across the continuum. This provides a direct connection between entanglement and experimentally accessible coincidence measurements. 

{\it Physical system:} We consider entanglement in photoionization with two ionic states $a$ and $b$ and one electron channel $\epsilon$, yielding states, $\ket{a,\epsilon}$ and $\ket{b,\epsilon}$, with coefficients, $\alpha(\epsilon)$ and $\beta(\epsilon)$, respectively. The final state wave function takes the generic form
\begin{equation}\label{eq:wf_prop}
\begin{split}
\ket{\Psi} 
    & = \int d\epsilon\, \alpha(\epsilon) \big[ \ket{a,\epsilon} + \eta(\epsilon) \,e^{i\phi(\epsilon)} \ket{b,\epsilon} \big], 
\end{split}
\end{equation}
where $\eta(\epsilon) = \abs{ {\beta(\epsilon)}/{\alpha(\epsilon)} }$ is the relative amplitude, and $\phi(\epsilon) = \arg\left[{\beta(\epsilon)}/{\alpha(\epsilon)}\right]$ is the relative phase of the composite states. 
One such scenario, generated through a shake-up process, is schematically shown in \cref{fig:2a}a). 
Here, the ionic states are entangled to the photoelectron energy due to the classical-like distinguishability of the distributions in the {\it preferred basis} for experiments, i.e. the energy eigenbasis $\ket{a}$ and $\ket{b}$ of the ion and kinetic energy of the electron
\cite{vedral_classical_2003,groisman_quantum_2005,xu_experimental_2010,modi_classical-quantum_2012}. We will refer to this as {\it amplitude entanglement} because the entanglement is independent of the quantum phase of the wave function, instead being governed by the relative amplitude, $\eta(\epsilon)$. The distinguishability of the photoelectrons allows amplitude entanglement to be detected via coincidence experiments with the ionic eigenstates \cite{stenquist_harnessing_2025,stenquist_entanglement_2025}. 
%
In contrast, {\it phase entanglement}, governed by the relative phase $\phi(\epsilon)$, between the ion and the electron can be generated via strong coupling of the ion, as illustrated in \cref{fig:2a}b) \cite{nandi_generation_2024}. Here, the two photoelectron peaks are entangled with the dressed states of the ion, {\it c.f.} Ref.~\cite{grobe_observation_1993}. The entanglement is due to the relative phase of the composite wave function. Hence, the entanglement can not be detected by direct measurement in the preferred basis.
The strong coupling of the ion requires non-perturbative, short-wavelength, ultrafast pulses, which can be generated at seeded free-electron-laser facilities, such as FERMI \cite{young_femtosecond_2010,allaria_highly_2012,fushitani_femtosecond_2016,prince_coherent_2016,rudenko_femtosecond_2017,young_roadmap_2018,lindroth_challenges_2019,kanter_unveiling_2011,nandi_observation_2022,maroju_attosecond_2023,dumergue_wave-packet_2024,li_imaging_2025}. Recently, improvements in pulse pairs delay stability \cite{gauthier_generation_2016,ardini_generation_2024} facilitate the study of phase-locked multi-pulse photoionisation and strong coupling with attosecond temporal precision.
We propose that such pulse sequences can be used to coherently control the generation and manifestation of entanglement via the phase of the pulses $\phi_I = \omega_0 t_I$, where $t_I$ is the time delay of pulse $I$ and $\omega_0$ is the resonant frequency. 
%
We investigate sequences of pulses as shown in \cref{fig:2a}c).
Such pulses are categorized in \cref{fig:2a}d), where, $+$ refers to pulses that have the same phase as the first pulse, $\phi_I=n2\pi$, and $-$ pulses that are out of phase with the first pulse, $\phi_I=(2n+1)\pi$. Fields are {\it phase continuous} if all pulses are in phase, i.e., all pulses share the same carrier wave. Otherwise, they are phase discontinuous. 
The resulting complex photoelectron spectra are analyzed by quantifying ion-electron entanglement, as well as the experimentally more accessible amplitude entanglement and classical correlation in the preferred basis. Finally, we discuss how entanglement is affected by decoherence due to the spatial separation of the electron states and temporal jitter of the pulses.  
Atomic units are used throughout, $e=\hbar=m_e=4\pi\epsilon_0=1$, unless otherwise stated.


\begin{figure*}
    \centering
    \includegraphics[width=0.9\linewidth]{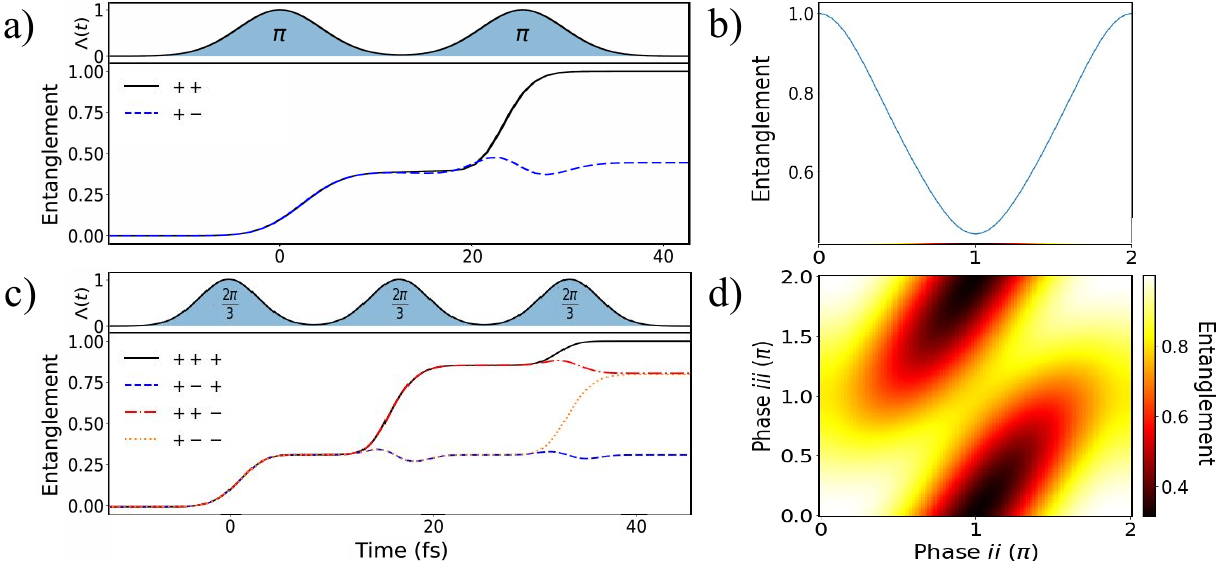}
    \caption{\textit{Electron-ion entanglement build-up for sequences of two and three short pulses.} a) Entanglement build-up for two in-phase $(++)$ and out-of-phase $(+-)$ $\pi$-pulses (illustrated at the top). b) Final entanglement in a), depending on their relative pulse-pair phase. c) Same as a) but for three pulses, each with area $2\pi/3$. d) Final entanglement in c) resolved over the phase of the second and the third pulse on the horizontal and vertical axes, respectively.  } 
    \label{fig:3}
\end{figure*}

\section{Results}\label{sec:results}


We will present results for helium atoms where ion-electron entanglement is generated after one-photon ionization, $\ket{1s^2}\rightarrow\ket{1s,\epsilon}$, through resonant strong coupling of the ion, $\ket{1s} \leftrightarrow \ket{2p}$, as schematically shown in \cref{fig:2a}b). For a single coherent pulse, this allows for the generation of a purely phase-entangled ion-electron system, while phase-locked pulse sequences, shown in \cref{fig:2a}c), can be used to coherently control the process and the manifestation of its quantum correlations. For details about the physical system, see Methods, \cref{Theory}. 

\subsection{Time-dependent Entanglement}\label{sec:res_entangle}

In this section, we present results for the build-up of entanglement between the ion and electron for pulse sequences, quantified using the von Neumann entropy 
\begin{equation}\label{eq:vN}
    S_\text{vN}(t) = - \Tr{\tilde \rho(t) \log_2[\tilde \rho(t)]} \in [0, \log_2(d)],
\end{equation}
where $\tilde \rho(t)$ is the time-dependent reduced density matrix of the ion or the electron conditioned on photoionization. 
Since full ion-electron entanglement can be generated when the ion is strongly coupled and driven through one Rabi cycle \cite{nandi_generation_2024}, we will consider pulse sequences where the absolute area adds up to $2\pi$. 

\subsubsection{Two $\pi$-pulses}\label{sec:twopulse}
In \cref{fig:3}a), we show the build-up of entanglement by a pair of pulses, each with area $\pi$. We find that if the pulse pair is in phase, $(++)$ black line, full entanglement is generated between the ion and the electron, $S_\text{vN} \approx 1$, in agreement with the single pulse case \cite{nandi_generation_2024}. Surprisingly, if the pulse pair is out of phase, $(+-)$ blue dashed line, we see that the second pulse does not induce any further entanglement, yielding the partial entanglement, $S_\text{vN} \approx 1/2$. In \cref{fig:3}b), the final ion-electron entanglement is resolved over the relative phase between the pulses. We observe the entanglement smoothly transition from its maximum value, when the pulses are in phase $(++)$, to its minimum, when the pulses are out of phase $(+-)$. 

\subsubsection{Three $\frac{2\pi}{3}$-pulses}\label{sec:threepulse}
In \cref{fig:3}c), we present the case of a field containing three pulses, 
each with area $2\pi/3$. We find that three pulses that are in phase, $(+++)$ black line, yield a fully entangled electron-ion system, as expected for a total $2\pi$ pulse area. Interestingly, we find that more entanglement is generated by the second pulse than by the other two pulses. 
In contrast, we observe that a pulse sequence where each pulse is out of phase with the previous pulse, $(+-+)$ blue dashed line, results in each subsequent pulse leaving the entanglement constant at the level of the first pulse. Hence, a pulse that is out of phase with the previous pulse will not yield further entanglement. 
Additionally, we find that both pulse sequences with a single phase discontinuity, $(++-)$ red dash-dotted line and $(+--)$ orange dotted line, generate equal amounts of entanglement. In the latter case, $(+--)$, where the second and third pulses are out of phase with the first, we find that it is the first and the third pulse that builds up the entanglement. In general, we find that a second out-of-phase pulse can generate further entanglement if the area exceeds that of the first pulse (equivalent to additional pulses). 

In \cref{fig:3}d) the final ion-electron entanglement in the three pulse case is resolved over the phase of the second pulse $\phi_{ii}$ and the phase of the third pulse $\phi_{iii}$. While the maximal entanglement is achieved when the pulses are in phase, corresponding to the four corners of \cref{fig:3}d), high entanglement is also observed at the saddle points $\phi_{iii}=\pi$, for $\phi_{ii}=0,\pi,2\pi$, corresponding to pulse sequences containing only one phase discontinuity. The lowest entanglement is found in the case where $\phi_{ii}=\pi$ and $\phi_{iii}=0$, corresponding to the case where the pulse sequence contains two complete phase discontinuities. Thus, a clear connection between the degree of entanglement and the number of phase discontinuities is found, with continuous variation for smooth phase changes in \cref{fig:3}b) and d). 

\begin{figure}
    \centering
    \includegraphics[width=1\linewidth]{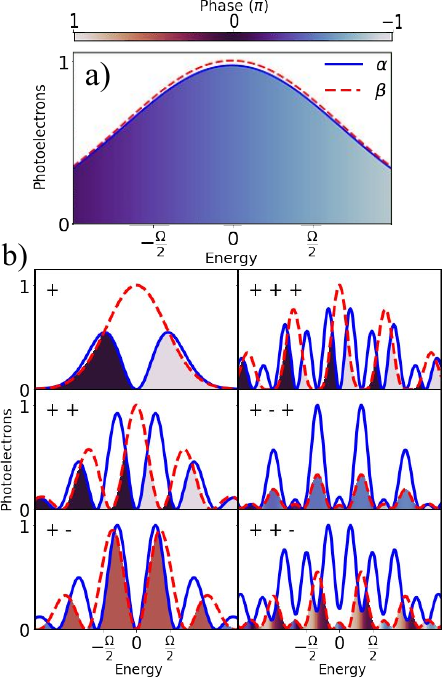}
    \caption{\textit{Phase-resolved photoelectron spectra from short Gaussian pulse sequences.} a) Photoelectron spectra for a single Gaussian pulse with $\pi$ area. b) Photoelectron spectra for pulse sequences with total absolute area $2\pi$, except for $(+)$, the corresponding entanglement dynamics are shown in \cref{fig:3}.}
    \label{fig:5}
\end{figure}

\subsection{Phase and Amplitude of Photoelectrons} \label{sec:nat_ent}
In \cref{fig:5} we will now present the complex electron-ion wavefunctions that correspond to the entanglement at the end of the pulse sequences in \cref{fig:3}. 
The photoelectron distributions of the ground and excited ionic state channels are $\rho_\alpha= \abs{\alpha(t,\epsilon)}^2$ and $\rho_\beta= \abs{\beta(t,\epsilon)}^2$, respectively. The relative phase is defined in \cref{eq:wf_prop}. 
The photoelectron relative energy scale is defined as $\epsilon=E^\text{kin} - [\omega_0-(\epsilon_{1s}-\epsilon_{1s^2})]$, see Method, \cref{Theory}.
In \cref{fig:5}a), we present the case of a single pulse with $\pi$ area, corresponding to the partially entangled system seen after the first pulse in \cref{fig:3}a). It is observed that neither channel has split into a doublet, which implies that the partial entanglement must arise from the smooth variation of phase over photoelectron energy. Physically, this is due to the large width of the photoelectron distributions originating from the two dressed states. They combine to create a single, wide, and phase-continuous photoelectron wave packet for each channel, see supplemental material \cite{Supplemental_Material}.

\begin{figure}[]
    \centering
    \includegraphics[width=0.95\linewidth]{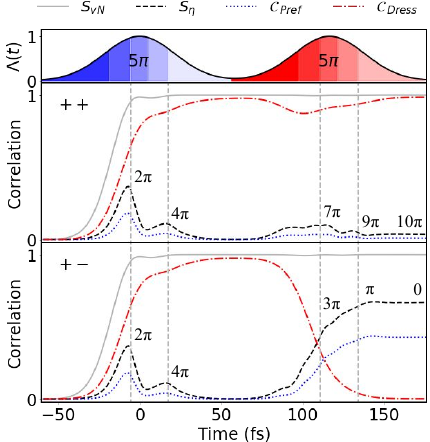}
    \caption{
    \textit{Generation of entanglement, amplitude entanglement and classical correlation by two long pulses.} Results are presented for in-phase $(++)$ and out-off-phase $(+-)$ pulse pairs. Total entanglement, grey lines, exhibit similar behaviour for both pulse sequences, while amplitude entanglement, black dashed lines, show contrasting behaviour during the second pulse. Classical correlation in the experimentally preferred basis of ionic states and photoelectron energy, blue dotted lines, exhibit similar behaviour as the amplitude entanglement. Classical correlation in the dressed state and photoelectron energy basis is shown in red dash dotted lines. Each pulse has $5\pi$ area. The pulse area is indicated by the vertical lines and the color under the envelope in the top panel. The pulse length is $\tau_I \approx 44$ fs and the separation is $\Delta t \approx 72.5$ fs. }
    \label{fig:4}
\end{figure}

In the panels of \cref{fig:5}b), photoelectron spectra are presented for pulse sequences with an absolute total area of $2\pi$. The single pulse $(+)$, with area $2\pi$, yields full entanglement. We observe that the ground state channel, $\alpha$, has a doublet shape, while the excited state, $\beta$, exhibits a single peak. This is in agreement with previous work, where it is shown that the ground state splits into a doublet around $2\pi$ and the excited state at around $3\pi$ pulse area \cite{yu_core-resonant_2018}. Here, we observe that the two channels partially overlap and that the phase of the two sides of the distribution differ by $\pi$, resulting in both phase and amplitude entanglement. This is in contrast to the distributions for long pulses, see \cite{nandi_generation_2024}, which are only phase entangled in the preferred basis. Thus, we have found that the duration of the pulse fundamentally changes the nature of the entanglement, as will be quantified in \cref{fig:4}. 

In the case of two pulses, corresponding to the entanglement dynamics in \cref{fig:3}a), we see that $(++)$ is phase and amplitude entangled due to the $\pi$ phase difference between the two sides of the electron spectra and the avoiding behaviour of the photoelectron peaks. For $(+-)$, however, we observe that the phase is mostly uniform throughout the distribution $\phi(\epsilon)\approx\pi/2$, yielding vanishing phase entanglement. However, the two channels are slightly shifted from each other in amplitude. We identify the general trend that odd fields generate amplitude entanglement. 
Interestingly, a second out-of-phase pulse does not make the photoelectron peaks narrower and, therefore, does not increase the entanglement. Instead, the second pulse rearranges the two channels $\ket{a,\epsilon}$ and $\ket{b,\epsilon}$ into having sine and cosine character, respectively, as shown in \cite{stenquist_harnessing_2025}. In this behaviour the Ramsey interference factor plays an important role, see supplemental material \cite{Supplemental_Material}.

Finally, for the case of three pulses, corresponding to the entanglement build-up in \cref{fig:3}c), we see a similar behaviour for $(+++)$ as for $(+)$ and $(++)$.  The observed antisymmetric phase distribution implies phase entanglement, while the avoiding behaviour of the amplitudes implies amplitude entanglement. We note that this unusual avoiding behaviour is due to the total $2\pi$ pulse area rather than the time symmetry. 
For $(+-+)$, the phase of the larger peaks is the same throughout the distribution, similarly to $(+-)$. However, the smaller peaks are inversely symmetric in phase, yielding a small degree of phase entanglement, see \cref{fig:3}c). Because the relative amplitude is approximately constant, $\eta(\epsilon) \approx \eta$, amplitude entanglement should be small. 
For $(++-)$, the photoelectron distributions are mostly avoiding, but the phase approximately differs by $\pi/2$, with respect to its mirrored component. This intricate behaviour yields a high, but not full, entanglement. 

The rather complicated phase and amplitude dependence of the wave packets, presented in \cref{fig:5}, is efficiently understood with the {\it basis-independent} measure of entanglement, 
presented in \cref{fig:3}. However, it does not provide any insight into the amplitude and phase nature of the entanglement, which depends on the choice of the basis.
%
In the next section, we show how amplitude entanglement, and classical correlations, develops in time and can be controlled without affecting the quantum entanglement.  

\section{Discussion}\label{sec:discussion}
We analyse how entanglement, amplitude entanglement and classical correlation depend on pulse area for different field symmetries in \cref{sec:Ent_Mes}. Additionally, we discuss the role of decoherence due to interaction with the environment in \cref{Sec:Decohere} and temporal decoherence in \cref{sec:Temp_Deco}. 

\subsection{Amplitude Entanglement}\label{sec:Ent_Mes}

In \cref{fig:2a}a), we presented a simple example of a system with amplitude entanglement in the preferres basis: one-photon ionization with a single shake-up channel. The helium atom in its ground state $\ket{1s^2}$ interacts with a high-frequency field that ionizes the atom, inducing transitions to the composite ion-electron ground state channel $\ket{1s,\epsilon_>}$
or via shake-up to the excited state channel $\ket{2s,\epsilon_<}$
, {\it c.f.} Ref.~\cite{ossiander_attosecond_2017}. For sufficiently long pulses, the photoelectron peaks $\epsilon_>$ and $\epsilon_<$ are non-overlapping, leading to vanishing coherences and to a diagonal reduced density matrix for the ion. The ion populations are $P_{1s} = \int d \epsilon \, |\alpha(\epsilon)|^2$ and $P_{2s} = \int d \epsilon \, |\beta(\epsilon)|^2$. Because the ion states are generally not equally populated, $P_{1s}\gg P_{2s}$, the ion and electron will not be fully entangled, $S_\text{vN}=-P_{1s}\log_2P_{1s}-P_{2s}\log_2P_{2s}< 1$. We refer to the final state as amplitude entangled because its entanglement can be described without relative phase dependence. 

Since populations can be observed in the preferred basis via coincidence measurements, we will now propose a lower bound for the quantum entanglement that is consistent with such populations. This amplitude entanglement, $S_\eta$ is constructed by calculating the von Neumann entropy of the reduced density matrix with a constant relative phase, {\it e.g.} \cref{eq:vN} with $\phi(\epsilon) = 0$ in \cref{eq:wf_prop}. We stress that the actual value of the constant phase does not affect the amplitude entanglement and that the coherences are not assumed to be zero.  
Defining a flat phase for the wave function in this way yields the maximal constructive interference of the coherences in the reduced density matrix, consequently $S_\eta\le S_\text{vN}$. If the ion channels are distinguishable in electron kinetic energy (traditional photoelectron spectroscopy case), the ionic coherences vanish, hence $S_\eta = S_\text{vN}$. 

Naturally, the amplitude entanglement can be compared to the classical correlation in the experimentally preferred basis. It is formally defined as the {\it classical mutual information},  which for the continuum electron $+$ qubit ion-system can be expressed as  
\begin{equation}
    \mathcal{C}_\text{Pref} 
    = \sum_{i\in \{a,b\} } \int d\epsilon \, P(i,\epsilon) \log\left[\frac{P(i,\epsilon)} {P(i)P(\epsilon)} \right],
    \label{eq:C}
\end{equation}
where we define the probabilities, $P(a,\epsilon) = |\alpha(\epsilon)|^2$, $P(a) = \int d\epsilon |\alpha(\epsilon)|^2$ (similar for $b$), and $P(\epsilon) = |\alpha(\epsilon)|^2 + |\beta(\epsilon)|^2$, see Method, \cref{Theory}. We note that in the case of shake-up the ion eigenstate and photoelectron kinetic energy is the optimal basis to resolve the correlations, ${\cal C}={\cal C}_\mathrm{max}=S_\mathrm{vN}$, {\it i.e.} traditional spectroscopy is a sufficient. If the pulse is sufficiently short to induce ionic coherences, however, the ion will have a higher purity \cite{pabst_decoherence_2011}, and the amount of entanglement and the classical correlations will be reduced. In this case, quantum tomography is required to map out the coherences of the quantum state, {\it c.f.} Ref.~\cite{laurell_measuring_2025}.   
%

In the case of photoionization with subsequent strong coupling, illustrated in \cref{fig:2a}b), classical correlation is not simply related to quantum entanglement, and traditional spectroscopy is not useful to resolve correlations. In \cref{fig:4}, we consider the correlation for in-phase and out-of-phase pulse-pairs, $(++)$ and $(+-)$, respectively, where each pulse has $5\pi$ pulse area. 
These pulses are long enough to individually create well defined photoelectron doublets corresponding to the two dressed states in the ion \cite{zhang_photoemission_2014,yu_core-resonant_2018}. 
As expected, we observe that the first pulse generates full entanglement, grey line, already at $\theta(t) \approx 2 \pi$ pulse area \cite{nandi_generation_2024}.  
%
After full ion-electron entanglement has been generated by the first pulse, the second pulse does not significantly change the degree of quantum entanglement. This is found to be generally true regardless of whether the pulse pair is in phase, $(++)$, or out of phase, $(+-)$. Similar results are found in cases containing additional pulses and other pulse delays. Once full entanglement has been generated, it cannot be significantly diminished by further resonant ionizing pulses.
%
While each pulse alone would not generate correlations in the preferred basis, their coherent addition was recently proposed to trigger strong correlations \cite{yu_core-resonant_2018,stenquist_harnessing_2025,stenquist_entanglement_2025}. In the following we will quantify these correlations, and more detailed ones, using formal measures of mutual information resolved over time.

In \cref{fig:4} we present the amplitude entanglement in black dashed lines and classical correlations in blue dotted lines. For the first pulse (same for $(++)$ and $(+-)$), we observe that the amplitude entanglement, $S_\eta$, increases later than the quantum entanglement, $S_\text{vN}$. The amplitude entanglement remains small for small pulse areas, since the relative strength of the two photoelectron channels is approximately constant, $\eta(\epsilon) \approx \eta$, analogous to the spectra in \cref{fig:5}a) for a single $\pi$-pulse. The amplitude entanglement then reaches a small peak around $2\pi$ pulse area. This is analogous to the photoelectron spectra presented in \cref{fig:5}b)$(+)$, where the ground state has split into a doublet, while the excited state remains a single peak. Subsequently, the excited state similarly splits, yielding overlapping electron spectra and significantly decreasing the amplitude entanglement. A faint amplitude entanglement peak is identified around $4\pi$ and attributed to bimaximas observed between the photoelectron doublet peaks. In this case the ground state has one bimaxima and the excited state has none, {\it c.f.} Refs.~\cite{zhang_photoemission_2014,yu_core-resonant_2018}. 
If the bimaximas in the two channels overlap, which happpens for the $5\pi$ pulse area, 
then we observe local minima in the amplitude entanglement. For a second in phase pulse, $(++)$, small modulations of the amplitude entanglement are induced by the same mechanism. 
%
%

In contrast, for $(+-)$, we observe that the second pulse induces strong amplitude entanglement, $S_\eta>50\%$, which is consistent with an ``avoiding behaviour'' of photoelectrons from different ion channels  \cite{yu_core-resonant_2018,stenquist_harnessing_2025,stenquist_entanglement_2025}. 
%
The classical correlation in the preferred basis, shown  in \cref{fig:4}, is observed to have remarkably similar behavior as the amplitude entanglement. 
In this way, the two quantities are closely related, but they differ in magnitude and interpretation. Unlike the classical correlations, amplitude entanglement is not computed from classical distributions --- rather it gives a minimal bound on quantum entanglement consistent with populations in a preferred basis.

Finally, we compute the classical correlation in the dressed state and photoelectron energy basis, $\mathcal{C}_\text{Dress}$, red dash-dotted line in Fig.~\ref{fig:4}. We observe that this correlation becomes large during the first pulse of the sequence after both channels split into photoelectron doublets, around $3\pi$. This is because each photoelectron peak, $\ket{\epsilon_<}$ and $\ket{\epsilon_>}$, becomes correlated with the corresponding dressed state $\ket{+}$ and $\ket{-}$, respectively. For the in-phase pulse pair $(++)$, this correlation is maintained during the second pulse and the correlation remains large. In contrast, for the out-of-phase case $(+-)$, the photoelectron peak structure is not maintained, which destroys the classical correlation between the dressed states and the photoelectron energy. 

\subsection{Role of Decoherence due to Spatial Separation} \label{Sec:Decohere}

Decoherence is typically induced when the system separates spatially. Electron trajectories with different energies, $\epsilon \neq \epsilon'$, naturally separate over time, $r(\epsilon)\neq r(\epsilon')$, and will be entangled to different parts of the environment \cite{zurek_decoherence_2003}.
%
As the environment is traced over the system becomes mixed, 
leading to decoherence as the off-diagonal elements of the density matrix gradually decay as the electrons become separated. We note that for a mixed system, the quantum correlation is not necessarily equal to the classical correlation, which therefore requires careful evaluation of the mutual information \cite{ollivier_quantum_2001,maziero_classical_2009}.

For the case of shake-up presented in \cref{fig:2a}(a) the full density matrix can be expressed via the ionic ground state connected to the higher photoelectron wave packet $\alpha(\epsilon_>)=\alpha^>$ and the excited state connected to the lower photoelectron wave packet $\beta(\epsilon_<)=\beta^<$ as
\begin{equation}
    \rho = \left[ 
    \begin{array}{cc}
        |\alpha^>|^2 & 
        \textcolor{lightgray}{\alpha^{>*}\beta^<} \\
        \textcolor{lightgray}{\beta^{<*}\alpha^>} 
        & |\beta^<|^2
    \end{array}
    \right] ,
\end{equation}
where the matrix elements in gray correspond to electron wave packets that separate in space, due to their different kinetic energies, centred around $\epsilon_>$ and $\epsilon_<$. After the ionization event, interaction with the environment will induce decoherence, acting on the spatially separated parts of the density matrix. Over time, the total density matrix would then become diagonal on the ionic blocks, meaning that we attain classical probability distributions \cite{zurek_decoherence_2003}. Since it gives the same reduced density matrix, decoherence does not affect the von Neumann entropy measure of entanglement despite the reduced quantumness of the evolving state. 

Analogously, the full density matrix in the case of ionization followed by strong coupling, illustrated in \cref{fig:2a}(b), can be expressed by considering the amplitude of the lower and higher photoelectron wave packet for the ground state $\alpha^<$ and $\alpha^>$, and the excited state $\beta^<$ and $\beta^>$ as
\begin{equation}
\rho = \left[
    \begin{array}{cccc}
        {\alpha^>}^*{\alpha^>} & \textcolor{lightgray}{{\alpha^>}^*{\alpha^<}} & 
        {\alpha^>}^*{\beta^>} & \textcolor{lightgray}{{\alpha^>}^*{\beta^<}} 
        \\
        \textcolor{lightgray}{{\alpha^<}^*{\alpha^>}} & {\alpha^<}^*{\alpha^<} & 
        \textcolor{lightgray}{{\alpha^<}^*{\beta^>}} & {\alpha^<}^*{\beta^<} 
        \\
        {\beta^>}^*{\alpha^>} & \textcolor{lightgray}{{\beta^>}^*{\alpha^<}} & 
        {\beta^>}^*{\beta^>} & \textcolor{lightgray}{{\beta^>}^*{\beta^<}} 
        \\
        \textcolor{lightgray}{{\beta^<}^*{\alpha^>}} & {\beta^<}^*{\alpha^<} & 
        \textcolor{lightgray}{{\beta^<}^*{\beta^>}} & {\beta^<}^*{\beta^<} 
    \end{array}
    \right],
\end{equation}
where the elements that depend on different central electron energies are presented in gray. Again, these elements disappear through decoherence as the electrons become spatially separated. 
From the total density matrix, the reduced density matrix is formed as 
\begin{equation}
    \tilde \rho= \! \int \! d \epsilon \left[
    \begin{array}{cc}
        {\alpha^>}^*{\alpha^>} + {\alpha^<}^*{\alpha^<}& 
        {\alpha^>}^*{\beta^>} + {\alpha^<}^*{\beta^<} 
        \\
        {\beta^>}^*{\alpha^>} + {\beta^<}^*{\alpha^<}& 
        {\beta^>}^*{\beta^>} + {\beta^<}^*{\beta^<} 
    \end{array}
    \right],
\end{equation}
which does not contain the elements that decohere. Hence, the mutual  information, which depends on the reduced density matrix, is unaffected in this case also.
Evidently, the coherences contain two terms, one with fast electrons and another with slow electrons. To evaluate their interference, we consider the wave function,   
\begin{equation}\label{Eq:WF_eps}\begin{split} 
    \ket{\Psi} 
    \!= \! \! \int \! d \epsilon \, \Big[
    |a\rangle\big( |\epsilon_<\rangle + |\epsilon_>\rangle\big) + 
    |b\rangle\big(|\epsilon_<\rangle-|\epsilon_>\rangle\big)\Big],
\end{split}\end{equation}
which reveals that the coherences cancel out, $\tilde \rho_{ab} = 0$ since ${\alpha^>}^*{\beta^>} = - {\alpha^<}^*{\beta^<}$. 
%
This is found to hold  in general for all resonant pulse sequences with sufficiently long pulse durations.
Because the coherences of the reduced density matrix are zero, the entanglement depends only on the diagonal elements. However, this does not imply that the system is amplitude entangled in the preferred basis, as it is the quantum phase of the system that induces the destructive interference of the coherences. Rather, the phase/amplitude nature of the entanglement depends on the wave function, not the density matrix. 

\begin{figure}
    \centering
    \includegraphics[width=1\linewidth]{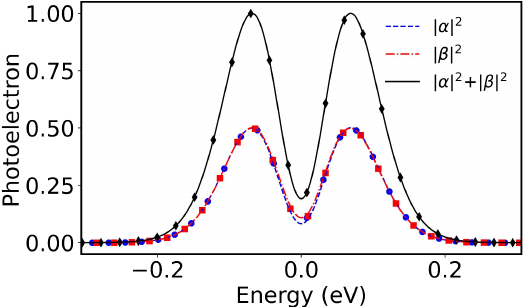}
    \caption{\textit{Photoelectron spectra generated by temporally-decoherent pulse pairs.} The photoelectron spectra, calculated by tracing over the relative phase, presented in lines is compared with single-pulse results presented in markers of the corresponding colour.}
    \label{fig:7}
\end{figure}

\subsection{Temporal Decoherence} \label{sec:Temp_Deco}

In contrast, temporal decoherence, generated by phase jitter between two pulses, will yield two peaks for each channel, $\alpha$ and $\beta$, as presented in \cref{fig:7} in blue dashed and red dot-dashed lines, respectively. The full photoelectron spectrum, $|\alpha|^2+|\beta|^2$, is presented in black lines. We note that the attained photoelectron spectra have the same shape as the single pulse spectra with the same properties as one of the pulses in the pair (markers). Thus, amplitude entanglement will only be revealed in the channel-resolved photoelectron spectra over multiple pulse shots when the pulse sequence is in a phase-locked configuration on the attosecond time scale.

\section{Conclusions}  

We have investigated the generation and control of entanglement in electron–ion pairs produced by photoionization in the strong-coupling regime. Using phase-locked pulse sequences, we demonstrate that entanglement can be coherently shaped: in-phase pulses enhance its buildup, while out-of-phase pulses can halt or delay its growth, leading to nontrivial dynamical control.

To characterize how these correlations manifest in experimentally accessible observables, we introduced a basis-resolved measure, termed amplitude entanglement, which identifies regimes where electron energy distributions are correlated with distinct ionic states. This provides a direct link between entanglement and coincidence measurements, such as joint detection of electron energy and ionic state or emitted photons.

We further show that while total entanglement cannot be reversed by subsequent driving, its observable structure can be strongly modified. Decoherence associated with spatially separated, non-overlapping electron wave packets has limited impact on the von Neumann entropy, whereas temporal phase coherence of the driving pulses is essential for maintaining control.

These results demonstrate that entanglement in photoionization is not only generated but can be actively shaped through coherent control, opening new possibilities for accessing and manipulating quantum correlations in ultrafast and strongly driven systems.
More generally, these findings highlight that in complex quantum systems the observable manifestation of entanglement depends not only on its magnitude but also on how it is distributed across measurement bases.

\section{Method}\label{Theory}

%
%
The energies of the helium atomic ground state, ionic ground state, and ionic excited state are: $\epsilon_{1s^2}=0$ eV, $\epsilon_{1s}=25.0$ eV, and $\epsilon_{2p}=65.8$ eV, respectively, and the transition matrix elements are $z_{1s,1s^2} = 0.502$ and $z_{2p,1s} = 0.373$. The pulses have Gaussian envelopes with the peak intensity $1.25\cdot10^{13} \;\text{Wcm}^{-2}$, yielding the maximum Rabi frequency $\Omega_0 = z_{2p,1s} E_0 = 0.19$ eV.
The central frequency $\omega_0 = 40.8$ eV.
The pure wave function is 
\begin{equation} \label{eq:wf}
    \ket{\Psi(t)} = g(t)\ket{g} + \int d\epsilon \, \big[\alpha(t, \epsilon) \ket{a, \epsilon} + \beta(t, \epsilon)\ket{b, \epsilon}\big],
\end{equation}
where $\ket{g}$ is the atomic ground state, $\ket{a, \epsilon}$ and $\ket{b, \epsilon}$ are the composite electron-ground and -excited ion state, respectively, with complex amplitudes: $g(t)$, $\alpha(t, \epsilon)$, and $\beta(t, \epsilon)$. 

\subsection{Quantum correlation}
The total correlation in the system is given by the quantum mutual information, using the von Neumann entropy \cref{eq:vN}, as 
\begin{equation}\label{eq:I_Q}
    I_Q = S_\text{vN}(\tilde \rho_\text{ion}) + S_\text{vN}( \tilde \rho_\epsilon) - S_\text{vN}(\rho),
\end{equation}
which for a pure state, $\rho$, is $I_Q = 2 S(\tilde \rho)$, where $\tilde \rho$ is the reduced density matrix of either subsystem \cite{henderson_classical_2001,zurek_decoherence_2003}. Consequently, the quantum and classical correlations $\mathcal{Q} =  \mathcal{C}_\text{max} = S_\text{vN}(\tilde\rho)$ . Furthermore, since all quantum correlation is entanglement for a pure state, the ion-electron entanglement can be quantified as $\mathcal{E} = S_\text{vN}(\tilde\rho)$, where 
\begin{equation}\label{eq:red_dens}
    \tilde \rho(t) = \int \! d  \epsilon \, \frac{1}{P_\alpha + P_\beta}
    \left[
    \begin{matrix}
        \abs{\alpha(t,\epsilon)}^2 & \alpha(t,\epsilon)^*\beta(t,\epsilon) \\
        \beta(t,\epsilon)^*\alpha(t,\epsilon)  & \abs{\beta(t,\epsilon)}^2
    \end{matrix} 
    \right]
\end{equation}
is the reduced post-measurement density matrix of the ion formed by conditioning the full density matrix on ionization \cite{haroche_exploring_2006} and renormalizing with the populations $P_\alpha = \int d\epsilon \, \abs{\alpha(t,\epsilon)}^2$ and $P_\beta = \int d\epsilon \, \abs{\beta(t,\epsilon)}^2$. 
%
%
%
The entanglement is independent of the basis representation.
%
%
%

The entanglement can be determined formally via Schmidt decomposition,
\begin{equation}
    |\Psi(t) \rangle_\text{ion,e} = \sum_i \sqrt{\lambda_i}\, |u_i\rangle_\text{ion} \otimes |v_i\rangle_\text{e},
\end{equation}
where $\ket{u_i}_\text{ion}$ and $\ket{v_i}_\text{e}$, are the Schmidt basis states, which are the eigenvectors of the reduced density matrix of the ion and electron, respectively. 
The Schmidt coefficients, $\lambda_i \in [0,1]$, are the eigenvalues of the reduced density matrix. If more than one Schmidt coefficient is non-zero, then the wave function must be expressed as a combination of Schmidt pairs $\{|u_i\rangle_\text{ion} \otimes |v_i\rangle_\text{e}\}$, and the system is entangled. The degree of entanglement can then be quantified using the von Neumann entropy as $S_\text{vN}=-\sum_i \lambda_i \log_2(\lambda_i)$. 

To analyse the nature of entanglement, the composite ion-electron wave function in \cref{eq:wf} can be reexpressed as \cref{eq:wf_prop}.
%
We refer to entanglement as {\it amplitude entanglement} if $\eta(\epsilon_1)\neq\eta(\epsilon_2)$ and {\it phase entanglement} if $\phi(\epsilon_1)\neq\phi(\epsilon_2)$, for $\epsilon_1 \neq \epsilon_2$. The system is maximally entangled when the electron components are orthogonal. The degree of amplitude entanglement $S_\eta$ can be quantified by computing the von Neumann entropy with a reduced density matrix with constant relative phase $\phi(\epsilon) = 0$ as
\begin{equation}\label{eq:red_dens__eta}
    \tilde \rho_\eta(t) = \int \! d  \epsilon \, \frac{1}{P_\alpha + P_\beta}
    \left[
    \begin{matrix}
        \abs{\alpha(t,\epsilon)}^2                         & \! \abs{\alpha(t,\epsilon)}\abs{\beta(t,\epsilon)} \\
        \abs{\beta(t,\epsilon)}\abs{\alpha(t,\epsilon)} \! & \abs{\beta(t,\epsilon)}^2
    \end{matrix} 
    \right].
\end{equation}

\subsection{Classical correlation}
The classical correlation is defined by the classical mutual information. For a bipartite system, expressed in the bases $X$ and $Y$, it is defined as
\begin{equation}
\mathcal{C} =  H(X) + H(Y) - H(X,Y),
\end{equation}
where the Shannon entropy 
\begin{equation}
H(X) = - \sum_x P(x) \log_2 P(x),
\end{equation}
depends on the probabilities of the states in basis $X$ (similar for $Y$ and $XY$) \cite{henderson_classical_2001,zurek_decoherence_2003}. The classical correlation in the ionic state and photoelectron energy basis is then given by $\mathcal{C} = I_C(\text{ion}:\epsilon)$.
The maximal classical correlation is the classical mutual information in the optimal (i.e. Schmidt) basis: $\mathcal{C}_\text{max} = \max_\text{basis}( I_C ) = S(\tilde \rho)$.

\subsection{Resonant strong coupling by a pulse sequence}\label{sec:strong_coupling}

We consider a linearly polarized $N$-pulse electric field 
\begin{equation}\label{Eq:Et}
    E(t) = \sum_I^N E_I \Lambda_I(t-t_I)\cos[\omega_0(t-t_I)],
\end{equation}
where $E_I$ is the electric field peak amplitude, $\Lambda_I(t) = \exp[-\ln(4)\tau_I^{-2}t^2]$ is the Gaussian envelope, 
with pulse duration $\tau_I$, centred on time $t_I$ for pulse $I = \{i,ii,iii,iv, ...\}$; having phase $\phi_{I} = \omega_0 t_I$. 
The atom-field interaction is $V(t) = z E(t)$, where $z$ is the position operator. 
The state amplitudes of a two-level system $a(t,t_0)$ and $b(t,t_0)$ can be computed using the area theorem, where the pulse area of pulse $I$ is $\theta_I(t,t_0) =  z_{ba} E_I\int_{t_0}^t dt'\, \Lambda_I(t'-t_I)$. An atom interacting with $N$ non-overlapping pulses evolves as $\ket{\Psi_N} = \hat U_N...\hat U_I...\hat U_i\ket{\Psi_0}$ where 
\begin{equation}\label{Eq:GenProp}
\hat U_I= \left[
\begin{array}{cc}
     \!\!\cos\left(\frac{\Theta_I}{2}\right) \!\!&\!\! -i\sin\left(\frac{\Theta_I}{2}\right) e^{-i\phi_I}\!\! \\ \\
     \!\!-i\sin\left(\frac{\Theta_I}{2}\right) e^{i\phi_I} \!\!&\!\! \cos\left(\frac{\Theta_I}{2}\right)\!\!
\end{array}\right]
\; , \; \ket{\Psi_0} = \left[
\begin{array}{c}
     1 \\ \\
     0
\end{array}\right],
\end{equation}
and $\Theta_I = \theta_I(\infty,-\infty)$. 
%
For sequences where the pulses are in or out of phase, this simplifies to  $a(t,t_0)=\cos[\theta(t,t_0)/2]$ and $b(t,t_0) = -i\sin[\theta(t,t_0)/2]$, where the total pulse area is defined as $\theta(t,t_0) = \sum_I \theta_I(t,t_0)$.
%
The area of each pulse can be calculated analytically for Gaussian pulses as
\begin{equation}
\begin{split}
    \theta_I(t,t_0) = \sqrt{\frac{\pi}{c}}\frac{z_{ba}E_I}{2}\Big\{
    & \text{erf}[\sqrt{c}(t  -t_I)]- \\
    & \text{erf}[\sqrt{c}(t_0-t_I)]\Big\}\cos(\phi_I), 
\end{split}
\end{equation}
where $\text{erf}$ denotes the error function.

%


\subsection{Composite ion-electron wave function}\label{sec:compWF}

The time-dependent amplitudes in \cref{eq:wf} are attained by solving for the terms of the Dyson-like expansion of the evolution $\ket{\Psi(t)} = U(t,t_0)\ket{g}$.
The zeroth-order wave function is attained by freely propagating the atomic ground state: $U_0(t,t_0)\ket{g} = g(t) \ket{g}$ with depletion:  
\begin{equation}\label{Eq:amp_g}
    \begin{split}
        g(t) =& \exp\left[ -\frac{\pi}{4} \int_{t_0}^{t} dt' \Omega_{ag}(t)^2 \right], 
    \end{split}
\end{equation}
for a flat continuum, following Yu and Madsen \cite{yu_core-resonant_2018}, where $\Omega_{ag}(t)=\sum_I z_{ag}E_I \Lambda_I(t - t_I)$. The first-order wave function is 
\begin{equation}
    \ket{\Psi^{(1)}(t)} = -i \int_{t_0}^t dt' U_R(t,t') V(t') U_0(t',t_0) \ket{g},
\end{equation}
where atom is ionized at time $t'\in[t_0,t]$. Subsequently, the ionic state is evolved by the Rabi propagator, with free evolution of the electron
\begin{equation}
\begin{split}
    \hat U_R(t,t') \ket{a, \epsilon} = 
    & a(t,t')\ket{a,\epsilon}e^{-i(\epsilon_a + E^\text{kin})(t-t')}  \\
    + & b(t,t')\ket{b,\epsilon}e^{-i(\epsilon_b + E^\text{kin})(t-t')},
\end{split}
\end{equation}
yielding the amplitudes in the rotating frame
\begin{equation}\label{Eq:GE-doub}
\begin{split}
    \alpha(t,\epsilon) &= \frac{z_{ag}}{i}\int_{t_0}^t dt' a(t,t') E(t') g(t') e^{i(\epsilon + \omega_0)  t'} \\
    \beta(t,\epsilon)  &= \frac{z_{ag}}{i}\int_{t_0}^t dt' b(t,t') E(t') g(t') e^{i(\epsilon  + \omega_0) t'}.
\end{split}
\end{equation}

\section{Funding} 
This work was supported by the Olle Engkvist Foundation: 194-0734; the Knut and Alice Wallenberg Foundation: 2019.0154, 2024.0212; and the Swedish Research Council: 2024-04247.

\section{Author Contributions}
A. Stenquist developed the model and produced the results under the supervision of J. M. Dahlström. Both authors analysed the results and wrote the manuscript. 

\section{Conflict of interest statement}
The authors declared no conflict of interest.

\bibliography{Train}

\end{document}